 \documentclass[conference, 10pt]{IEEEtran}

\usepackage{fullpage}
\usepackage{citesort}

\usepackage{epsfig}
\usepackage{graphicx}
\usepackage{caption}
\usepackage{subcaption}
\usepackage{fancybox}
\usepackage{color}
\usepackage{amssymb}
\usepackage{color}

\usepackage[cmex10]{amsmath}
\usepackage{amsfonts}
\usepackage{cases}
\usepackage{url}

\usepackage{epstopdf}

\newtheorem{theorem}{Theorem}

\newcommand{\cl}[1]{{\cal #1}}

\title{\vspace*{-5mm} A Parallel Two-Pass
MDL Context Tree Algorithm
for Universal Source Coding \vspace*{-3mm}}

\author{\IEEEauthorblockN{Nikhil Krishnan,\IEEEauthorrefmark{1}
Dror Baron,\IEEEauthorrefmark{1} and Mehmet K{\i}van\c{c} M{\i}h\c{c}ak }
\IEEEauthorblockA{\IEEEauthorrefmark{1}Department of Electrical and Computer Engineering\\
North Carolina State University;
Raleigh, NC 27695, USA\\
Email: $\lbrace $nkrishn, barondror$\rbrace $@ncsu.edu, kivancmihcak@gmail.com \vspace*{-5mm}}
}

\begin{document}

\thanks{This work was supported in part by the National Science Foundation under Grant CCF-1217749 and in part by the U.S. Army Research Office under Grant W911NF-04-D-0003.}

\maketitle

\begin{abstract}
We present a novel lossless universal source coding algorithm
that uses parallel computational units 
to increase the throughput. The length-$N$
input sequence is partitioned into $B$ blocks.
Processing each block independently of the other blocks
can accelerate the computation by a factor of $B$,
but degrades the compression quality.
Instead, our approach is to first estimate the
minimum description length (MDL)
source  underlying the entire input,
and then encode each of the $B$ blocks in parallel
based on the MDL source.
With this two-pass approach, the compression loss
incurred by using more parallel
units is insignificant.
Our algorithm is work-efficient, i.e.,
its computational complexity is $O(N/B)$.
Its redundancy is approximately
$B\log(N/B)$ bits above Rissanen's lower bound
on universal coding performance, with respect to any
tree source whose maximal depth is at most $\log(N/B)$.
\end{abstract}


%
\IEEEpeerreviewmaketitle

\section{Introduction} \label{sec:intro}
\vspace*{-2mm}
\subsection{Motivation} \label{subsec:intro:motivation}

With the advent of cloud computing and big data problems, the amount of data processed by computer and communication systems has increased rapidly.
This growth necessitates the use of efficient and fast compression algorithms to comply with data storage and network bandwidth requirements. 
At present, typical lossless data compression algorithms,
which are implemented in software, run at least an order of magnitude
slower than the throughput delivered by hard disks; they are even slower when compared to optical communication devices.
Therefore, lossless compression may be a computational bottleneck.

One obvious approach to speed up compression algorithms
is to implement them in special-purpose hardware~\cite{ArmingIBWT}. Although hardware implementation
may accelerate compression by approximately an order of
magnitude, there are still many systems where this does
not suffice. Ultimately, in order for lossless compression
to become appealing for a broader range of applications,
we must concentrate
more on efficient new algorithms.

Parallelization is a possible direction for fast source coding
algorithms. By compressing in parallel, we may obtain
algorithms that are faster by orders of magnitude.
However, with a naive parallel algorithm, which
consists of partitioning the original input into $B$ blocks and
processing each block independently of the other blocks,
increasing $B$ degrades the compression quality~\cite{FRT96}. Therefore,
naive parallel compression has limited potential.
Sharing information across blocks can improve the compression quality of data~\cite{Beirami2012}.

\subsection{Related work}

Stassen and Tjalkens~\cite{ST2001} proposed a
parallel compression algorithm based on context tree
weighting~\cite{Willems1995CTW} (CTW),
where a common finite state machine (FSM)
determines for each symbol
which processor should process it. Since the FSM processes the
original length-$N$ input in $O(N)$ time, Stassen and Tjalkens' method
does not support scalable data rates.

Franaszek et al.~\cite{FRT96} proposed a
parallel compression algorithm, which is related to LZ77~\cite{Cover06}, where the construction of a dictionary is
divided between multiple processors.
Unfortunately, the redundancy
(excess coding length above the entropy rate) of LZ77 is high.

Finally, Willems~\cite{Willems2000} proposed
a variant of CTW with $O(ND/B)$ time complexity, where $D$
is the maximal context depth that is processed.
Unfortunately, Willems' approach will not compress as well as CTW,
because probability estimates will be based on
partial information in between synchronizations
of the context trees.

\subsection{Contributions}

This paper presents a novel minimum description length~\cite{Rissanen1978} (MDL) source  coding algorithm that coordinates
multiple computational units running in parallel,
such that the compression loss incurred by
using more computational units is insignificant.
Our main contributions are ({\em i}) our algorithm is {\em work-efficient}~\cite{CLR}, i.e., it compresses $B$ length-($N/B$) blocks in
parallel with $O(N/B)$ time complexity, and
({\em ii}) the redundancy of our algorithm is approximately $B\log(N/B)$ bits above the lower bounds on the best
achievable redundancy.

The remainder of the paper is organized as follows. We review preliminary material in 
Section~\ref{sec:prelim} 
and propose our new parallel two-pass MDL algorithm in Section~\ref{sec:basic}. 
Finally, Section~\ref{sec:num_res} discusses numerical
results.

\section{Source coding Preliminaries} \label{sec:prelim}

\subsection{Universal Source coding} \label{subsubsec:universal}

Lower bounds on the redundancy serve as benchmarks for compression quality. 
Consider length-$N$  sequences $x$ generated by a stationary ergodic source over a finite alphabet $\cl{X}$, i.e. $x\in\cl{X}^N$.
For an individual sequence $x$, the {\em pointwise redundancy} with respect to (w.r.t.) a class $\cl{C}$ of source models is
\begin{equation*}
\vspace*{-2mm}
\rho(x)\triangleq{l(x)-N\widehat{H_x}},
\end{equation*}
where $l(x)$ is the length of a uniquely decodable code~\cite{Cover06} for $x$,
and  $\widehat{H_x}$ is the entropy of $x$ w.r.t. the best model in $\cl{C}$
with parameters set to their maximum likelihood  (ML) estimates. Weinberger et al.~\cite{WMF1994} proved for a source with $K$ (unknown) parameters that
\begin{equation}
\vspace*{-2mm}
\label{eq:WMF94}
\rho(x)  \geq \frac{K}{2}(1-\epsilon)\log(N),
\end{equation}
where $\log(\cdot )$ denotes the base-2 logarithm, and for any $\epsilon>0$,
except for a set of inputs whose probability vanishes
as $N\rightarrow\infty$.
Similarly, Rissanen~\cite{Rissanen1996} proved that, for universal
coding of independent and identically distributed (i.i.d.) sequences, the worst case redundancy (WCR) is at least
$\frac{|\cl{X}|-1}{2}\log(N)+C_{|\cl{X}|}+o(1)$
bits, where $|\cl{X}|$ denotes cardinality of $\cl{X}$, and $C_{|\cl{X}|}$ was specified.
Because i.i.d. models are too simplistic for modeling
``real-world" inputs, we use tree sources instead.

\subsection{Tree sources}

\begin{figure}[t]
\begin{center}
\includegraphics[angle=0,width=5.5 cm]{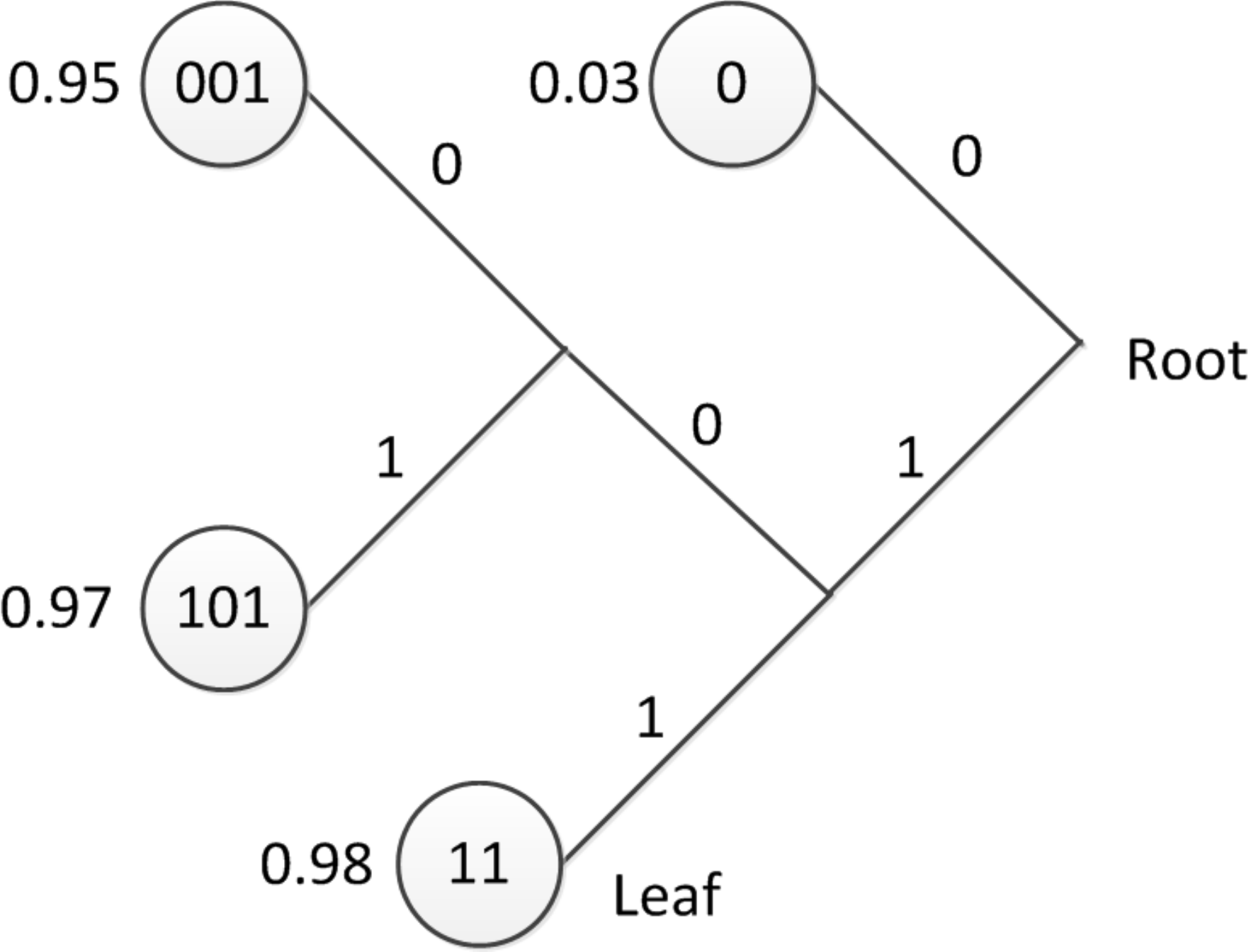}
\end{center}
\vspace*{-2mm}
\caption
{A tree source over $\cl{X}=\{0,1\}$.
The states are $\cl{S}=\{0,11, 001, 101\}$
and the conditional probabilities are $p(x_i=1|0)=0.03$,
$p(x_i=1|11)=0.98$, $p(x_i=1|001)=0.95$, and $p(x_i=1|101)=0.97$.}
\vspace*{-5mm}
\label{fig:tree}
\end{figure}

Let $x_i^j$ denote the {\em sequence}
$x_i,x_{i+1},\ldots,x_j$  where $x_k\in\cl{X}$
for $i\leq{k}\leq{j}$.
Let $\cl{X}^*$ denote the set of finite-length
sequences over $\cl{X}$.
Define a {\em context tree source}  $\{\cl{S},\Theta\}$~\cite{Willems1995CTW}
as a finite set of sequences called states
$\cl{S}\subset\cl{X}^*$ that is
complete and proper~\cite[p.654]{Willems1995CTW},
and a set of conditional probabilities $\Theta=\{p(\alpha|s):\ \alpha\in\cl{X},\ s\in\cl{S}\}$.
We say that $s$ {\em generates} symbols following it.
Because $\cl{S}$ is complete and proper, the sequences
of $\cl{S}$ can be arranged as leaves on
an $|\cl{X}|$-ary tree~\cite{CLR}
(Fig.~\ref{fig:tree});  the
unique state $s$ that generated
$x_i$ can be determined by entering the tree at the root,
first choosing branch $x_{i-1}$, then branch $x_{i-2}$,
and so on, until some leaf $s$ is encountered.
Let $D\triangleq\max_{s\in\cl{S}}|s|$ be the {\em maximum context depth}.
Then the string
$x_{i-D}^{i-1}$ uniquely determines the current state $s$;
the previous symbols $x_{i-L}^{i-1}$ ($L\leq{D}$)
that uniquely determine the current state $s$ are called the
{\em context}, and $L$ is called the {\em context depth}
for state $s$.

\subsection{Semi-predictive and two-pass source coding}

Consider a tree source structure $\cl{S}$ whose explicit
description requires $l_{\cl{S}}$ bits,
and denote the probability of the input sequence $x$
conditioned on the tree source structure
$\cl{S}$ by $p_{\cl{S}}(x)$.
Using $\cl{S}$, the coding length required
for $x$ is $l_{\cl{S}}-\log(p_{\cl{S}}(x))$.
Define the {\em MDL tree source structure} $\cl{\widehat{S}}$
as the tree source structure that provides the
shortest description of the data, i.e.,
\begin{equation*}
\vspace*{-2mm}
\cl{\widehat{S}} \triangleq
\arg\min_{\cl{S}\in\cl{C}}
\left\{l_{\cl{S}}-\log(p_{\cl{S}}(x))\right\},
\end{equation*}
where $\cl{C}$ is the class of tree source models
being considered.
The {\em semi-predictive} approach~\cite{BWTMDL,Volf1995CTWMDL,WillemsCISS}
processes the input $x$
in two phases.
{\em Phase~I} first estimates $\cl{\widehat{S}}$
by context tree pruning (CTP), which is a form of dynamic
programming for coding length minimization
(c.f. Baron~\cite{BaronThesis}
for details). 
The structure of $\widehat{\cl{S}}$ is then encoded explicitly.
{\em Phase~II} uses $\widehat{\cl{S}}$ to encode the
sequence $x$ sequentially, where the parameters $\widehat{\Theta}$ are estimated while encoding $x$.
The {\em decoder} first determines
$\widehat{\cl{S}}$, and afterwards uses it to decode $x$ sequentially.

Two-pass MDL codes for tree sources describe both
$\widehat{\cl{S}}$ and $\widehat{\Theta}$ in Phase~I using CTP, and encode $x$
in Phase~II. 
We use a two-pass approach instead of
a semi-predictive approach,
because estimating $B$ sets of parameters in parallel,
one for encoding each of the $B$ blocks in Phase~II,
has $ \rho(x) \approx 0.5B|\cl{S}|\log(N/B)$, whereas the two-pass approach has $\rho(x) \approx 0.5|S|\log(N)$, and the latter redundancy is smaller.

\section{Proposed Algorithm} \label{sec:basic}

We present a new {\em Parallel Two-Pass MDL} (PTP-MDL) algorithm. In order to keep the presentation simple, we restrict
our attention to a binary alphabet, i.e., $\cl{X}=\{0,1\}$;
the generalization to non-binary alphabets is
straightforward.
We will show that PTP-MDL has $O(N/B)$ time
complexity when we restrict $D \leq \log (N/B)$, while still approaching the pointwise redundancy bound (\ref{eq:WMF94}).
This enables scalable data rates without
a factor-$B$ increase in the redundancy.

\subsection{Overview}

\begin{figure}[t]
\hspace*{-1mm}
\includegraphics[angle=0,width=8 cm]{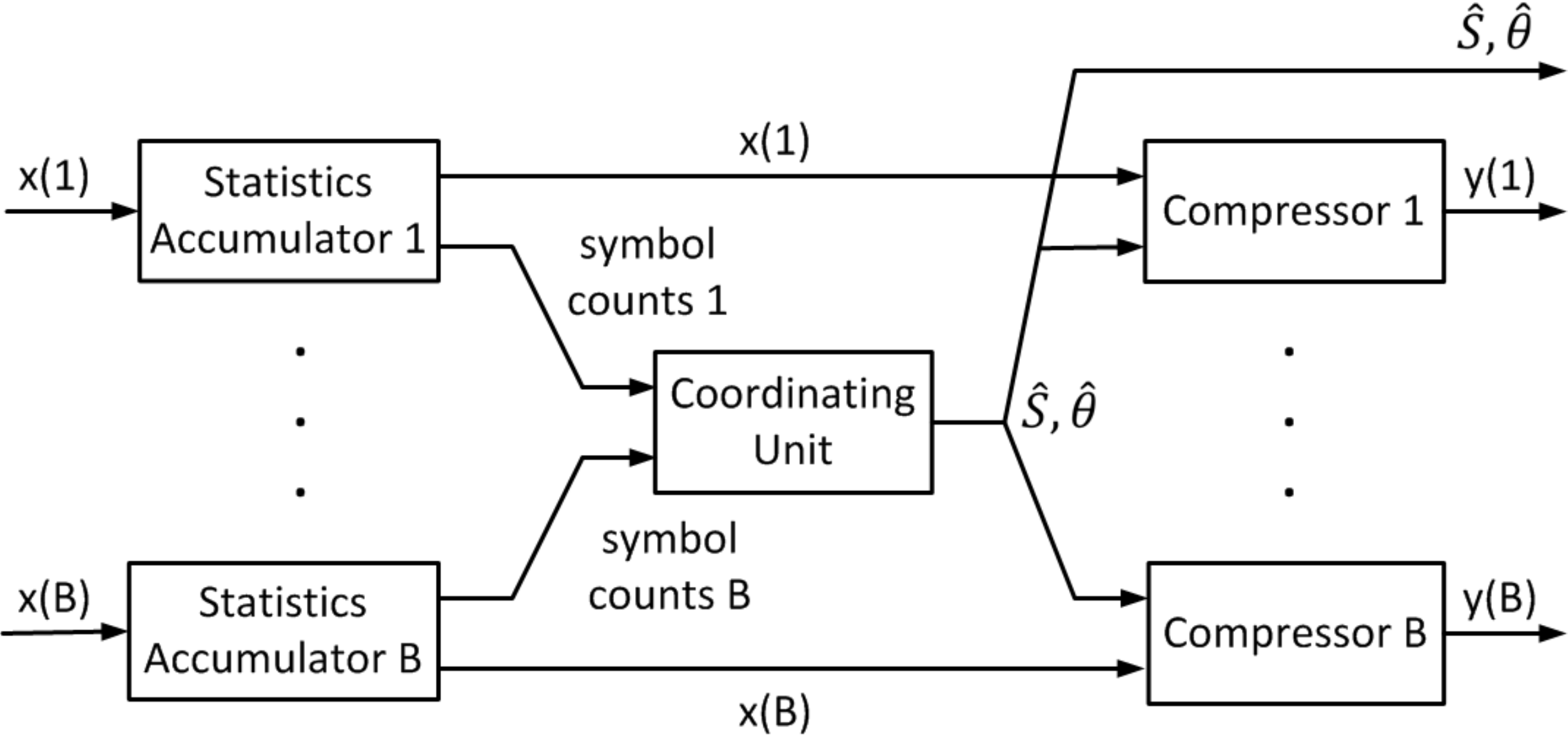}
\vspace*{-1mm}
\caption
{Block diagram of the PTP-MDL encoder.}
\vspace*{-5mm}
\label{fig:PTP-MDL}
\end{figure}

A block diagram of a possible implementation of the PTP-MDL encoder
is shown in Fig.~\ref{fig:PTP-MDL}.
In Phase~I, the PTP-MDL encoder employs $B$ computational units called {\em parallel units} (PUs)
that work in parallel to accumulate statistical information
on $B$ blocks in $O(N/B)$ time, and a
{\em coordinating unit} (CU) that controls the PUs and
computes the MDL source estimate $\{\widehat{\cl{S}},\widehat{\Theta}\}$.

Without loss of generality, we assume $N/B \in {\mathbb Z}^+$.
Define the $B$ blocks as
$x(1)=x_1^{N/B},
x(2)=x_{N/B+1}^{2N/B},
\ldots,
x(B)=x_{N-N/B+1}^{N}$.
PU~$b$, where $b\in\{1,\ldots,B\}$,
first computes for each depth-$D$ context $s$
the {\em block symbol counts} $n^{\alpha}_s(b)$,
which are the number of times 
$\alpha$ is
 generated by $s$ in $x(b)$,
\begin{equation*}
n^{\alpha}_s(b) \triangleq
\sum_{i=(b-1)(N/B)+ D+1}^{b(N/B)} 1_{\left\{x_{i-D}^{i}=s\alpha\right\}},
\alpha\in\cl{X},
\label{eq:def:n_s_alpha_b}
\end{equation*}
where $s\alpha$ denotes concatenation of $s$ and $\alpha$.
For each state $s$ such that $|s|<D$, the CU
either retains the children states $0s$ and $1s$ in the
MDL source, or prunes them and only retains $s$,
whichever results in a shorter coding length.
Details of the pruning decision appear in Section \ref{subsec:basic:phaseI:MDL}.
Note that the serial MDL source considers the last $D$ symbols from the previous block as context for counting the first $D$ symbols of the current block (except the first block). However, using the serial MDL source is suboptimal in PTP-MDL, because this source does not reflect the actual symbols compressed by PTP-MDL.

In Phase~II, each of the $B$ blocks is compressed by a PU.
For each symbol $x_i(b)$, PU~$b$ first determines the generator state $G_i(b)$,
the state $s$ that generated the symbol $x_i(b)$.
PU~$b$ then assigns $x_i(b)$ a probability
according to the parameters that were
estimated by the CU in Phase~I, and sequentially feeds the
probability assignments to an arithmetic encoder~\cite{Cover06}.

The structure of the decoder is similar to that of Phase~II.
The approximated MDL source structure $\widehat{\cl{S}}$
and quantized parameters $\widehat{\Theta}$ are
first derived from the parallel source description
(see Section~\ref{subsec:basic:bitstream}).
Then, the $B$ blocks are decompressed by $B$ decoding blocks.
In decoding block~$b$, each symbol $x_i(b)$ is sequentially
decoded by determining $G_i(b)$, assigning
a probability to $x_i(b)$ based on the parameter
estimates, and applying an arithmetic decoder~\cite{Cover06}.

\subsection{Parallel source description} \label{subsec:basic:bitstream}

\subsubsection{Two-part codes in the PTP-MDL algorithm}
\label{subsubsec:twopart}

Having received the block symbol counts $n^{\alpha}_s(b)$  from the PUs, the CU computes the {\em symbol counts} generated by state $s$ in the entire sequence $x$,
\begin{equation}
\vspace*{-2mm}
n^{\alpha}_s = \sum_{b=1}^B n^{\alpha}_s(b),\qquad\alpha\in\cl{X}.
\label{eq:n^alpha_s}
\end{equation}
The CU can then compute the ML parameter estimates
of $p(1|s)$ and $p(0|s)$,
\begin{equation*}
\vspace*{-2mm}
\theta_s\triangleq \theta_s^1 = \frac{n_s^1}{n_s^0+n_s^1}
\quad \quad
\text{ and }
\quad \quad
\theta_s^0 = 1 - \theta_s^1,
\end{equation*}
respectively. 
The ML parameter estimates for each state $s$ are quantized into one of

\begin{equation}
K_s  \triangleq  \left\lceil\sqrt{
2\pi^2\ln(2)\left(\frac{1}{2}-\frac{3}{16\ln(2)}\right)N}\right\rceil
 \approx \left\lceil 1.772\sqrt{N}\right\rceil
\label{eq:K^*}
\end{equation}
representation levels based on Jeffreys' prior~\cite{BaronBreslerMihcak2003}, where $\lceil \cdot \rceil$ denotes rounding up.
The representation levels and bin edges are computed using a Lloyd-Max procedure~\cite{BaronBreslerMihcak2003}.
The bin index and representation level for state $s$ are denoted by $k_s$ and $r_s$, respectively.
Denoting the quantized ML estimate of $\theta_s^{\alpha}$
by $\widehat{\theta}_s^{\alpha}$, we have $\widehat\theta_s^1=r_s$
and $\widehat\theta_s^0=1-r_s$. Recall that,
at the end of Phase~I, the CU
has computed the MDL structure estimate $\widehat{\cl{S}}$.
If $s\in\widehat{\cl{S}}$, then the first part of the two-part code
for symbols generated by $s$
consists of encoding $k_s$ with $\log(K_s)$ bits.
The WCR using this quantization approach is 1.047 bits per state above Rissanen's redundancy bound~\cite{Rissanen1996,BaronBreslerMihcak2003}.

In Phase~II, which implements the second part of
the two-part code, each PU~$b$ encodes its
block $x(b)$ sequentially. For each symbol $x_i(b)$,
PU~$b$ determines $G_i(b)$.
The symbol $x_i(b)$
is encoded according to the probability assignment
$\widehat{p}(x_i(b))\triangleq\widehat{\theta}_{G_i(b)}^{x_i(b)}$
with an arithmetic encoder~\cite{Cover06}.
Thus, the probability assigned by all $B$ PUs
to the symbols in $x$ whose generator state is $s$ is

\begin{equation}
\prod_{b=1}^B
\prod_{\left\{i:\ G_i(b)=s,\ i>D\right\}}
\widehat{p}(x_i(b)) =
(r_s)^{n_s^1}(1-r_s)^{n^0_s}.
\label{eq:length_part_two}
\end{equation}
Equation (\ref{eq:length_part_two})  provides the same redundancy for two-part codes in a
parallel compression system as we would obtain in a serial
system~\cite{BaronThesis}.

\subsubsection{Coding lengths in Phases~I and~II}
\label{subsec:protocol}

In Phase~I, the structure $\widehat{\cl{S}}$ is described with
the {\em natural code}~\cite{Willems1995CTW}.
 For a binary alphabet,
$|\text{natural}_{\widehat{\cl{S}}}|\leq{2}|\widehat{\cl{S}}|-1$ bits;
this is the {\em model redundancy} of PTP-MDL.
The parameters $\widehat{\Theta}$ are described as the $|\widehat{\cl{S}}|$ indices $k_s$ in the order in
which the leaves of $\widehat{\cl{S}}$ are reached in a depth-first
search~\cite{CLR}; this description can be implemented with
arithmetic coding~\cite{Cover06}.
The corresponding coding length
is the {\em parameter redundancy} of PTP-MDL.
We denote the length of the descriptions of $\widehat{\cl{S}}$
and $\widehat{\Theta}$ generated in Phase~I
by $l^I_{\cl{S}}$ bits. Using (\ref{eq:K^*}),
\vspace*{-2mm}
\begin{eqnarray}
l^I_{\cl{S}} &=&
|\text{natural}_{\cl{S}}|
+|\cl{S}|\log(K_s) \label{eqn:l1} \\
&\lessapprox&
\left[2|\cl{S}|-1\right]
+|\cl{S}|
\left[\log(1.772)+\frac{1}{2}\log(N)\right]. \nonumber
\end{eqnarray}
\vspace*{-1mm}

In Phase~II, the coding length is mainly determined by symbol probabilities conditioned on generator states as given by (\ref{eq:length_part_two}).
There are two additional terms that affect the coding
length in Phase~II.
First, {\em coding redundancy} for each arithmetic encoder  with $\log(N)$
bits of precision requires $O(1) \leq 2$ bits~\cite{Cover06}.
Second, {\em symbols with unknown context}  at the beginning of $x(b)$; 
we encode the first $D$ symbols of each block $x(b)$ directly using $D$ bits per block.
Denoting the combined length of all $B$ codes
in Phase~II by $l^{II}_{\cl{S}}$ bits, we have
\begin{equation}
\vspace*{-2mm}
l^{II}_{\cl{S}}
\lessapprox
B\cdot{(D+2)}
-\sum_{s\in\cl{S}}
[n_s^1 \log(r_s) + n_s^0 \log(1-r_s)]
\label{eq:length:PhaseII}.
\end{equation}
Combining (\ref{eqn:l1}) and (\ref{eq:length:PhaseII}), we have the following result for the redundancy.
\begin{theorem}~\cite{BaronThesis}
The pointwise redundancy of the PTP-MDL algorithm over the ML entropy of the input sequence x w.r.t. the MDL souce structure $\widehat{S}$  satisfies
\begin{equation*}
\rho(x) < B \left[ \log \left( \frac{N}{B}\right)  + 2  \right] + \frac{|\widehat{S}|}{2} \left[ \log \left(N \right) + O\left(1\right) \right].
\end{equation*}
\end{theorem}

Note that the redundancy for naive parallel compression is upper bounded by $ B \left[ \log \left( \frac{N}{B}\right)  + 2 + \frac{|\widehat{S_n}|}{2} \left[ \log \left(N \right) + O\left(1\right) \right] \right]$, where $\widehat{S_n}$ is the estimated tree structure with the largest number of states among the $B$ tree structures.

\subsection{Phase~I} \label{subsec:basic:phaseI}

\subsubsection{Computing block symbol counts}
\label{subsec:basic:phaseI:bscounts}

Computational unit~$b$ computes $n^{\alpha}_s(b)$
for all $2^{D}$ depth-$D$ leaf contexts $s$.
In order for PU~$b$ to compute all block symbol counts in
$O(N/B)$ time, we define the {\em context index} $c_i(b)$
of the symbol $x_i(b)$ as
\begin{equation}
c_i(b) \triangleq
\sum_{j=0}^{D-1} 2^j x_{j+i-D}(b),
\label{eq:def:c_i(b)}
\end{equation}
where $i\in\{D+1,\ldots,N/B\}$ and
$x_{j+i-D}(b)\in\{0,1\}$,
hence $c_i(b)\in\{0,\ldots,2^{D}-1\}$.
Note that $c_i(b)$ is the binary number represented by
the context $s=x_{i-D}^{i-1}(b)$. Hence, it can
be used as a pointer to the address containing the
block symbol count $n_s^{\alpha}(b)$ for $s=x_{i-D}^{i-1}(b)$.
Moreover, the property
\begin{equation}
c_{i+1}(b)=
\frac{c_i(b)}{2}+
2^{D-1}x_i(b)-
\frac{x_{i-D}(b)}{2}
\label{eq:property:update_c_i}
\end{equation}
enables the computation of
all $N/B-D$ context indices of the symbols of $x(b)$ in $O(N/B)$
time complexity.

\subsubsection{Constructing context trees}
\label{subsec:basic:phaseI:counts}

Because we restrict our attention to
depth-$D$ contexts, it suffices for PU~$b$ to compute
$\{n^{\alpha}_s(b)\}_{\alpha\in\cl{X},\ s\in\cl{X}^{D}}$,
all the block symbol counts of all the leaf contexts
of a full depth-$D$ context tree.
Information on internal nodes of the context
tree, whose depth is less than $D$, is computed
from  the block symbol counts of the leaf contexts.

If $|s|=D$, then the CU gets
$\{n^{\alpha}_s(b)\}_{\alpha\in\cl{X}}$
from the PUs and computes $n_s^{\alpha}$
with (\ref{eq:n^alpha_s}).
Alternatively, $|s|<D$, the CU
recursively derives $n_s^{\alpha}$
by adding up the symbol counts of children states, i.e.,
\begin{equation}
\label{eq:n^s_children}
n_s^{\alpha}= n_{0s}^{\alpha} + n_{1s}^{\alpha},
\qquad\forall\alpha\in\cl{X}.
\end{equation}

\subsubsection{Computing the MDL source $\{\widehat{\cl{S}},\widehat{\Theta}\}$}
\label{subsec:basic:phaseI:MDL}

For each state $s$, we either retain the children states
$0s$ and $1s$ in the tree or merge them into a single
state, according to which decision minimizes the coding length.
The coding length $l_s$ of the two-part code that
describes the symbols generated by $s$ is
\begin{equation}
l_s =
\overbrace{\log(K_s)}^{\text{Part~I}}
\overbrace{-n^0_s\log(1-r_s)-n^1_s\log(r_s)}^{\text{Part~II}}.
\label{eq:l_s:binary}
\end{equation}

We now derive the coding length required for state $s$,
which is denoted by $\text{MDL}_s$.
For $|s|=D$, $n_s^0$ and $n_s^1$
are computed with (\ref{eq:n^alpha_s}),
$l_s$ is computed with (\ref{eq:l_s:binary}),
and $\text{MDL}_s=l_s$.
For $|s|<D$, we compute $n_s^\alpha$
hierarchically with (\ref{eq:n^s_children}),
after already having processed the children states.
In order to decide whether to prune the tree,
we compare $\text{MDL}_{0s}+\text{MDL}_{1s}$ with $l_s$.
Because retaining an internal node requires
the natural code~\cite{Willems1995CTW}
to describe that node (with $1$ bit),
\begin{equation*}
\text{MDL}_s=
\left\{
\begin{matrix}
l_s \quad \text{if $|s|=D$} \\
1+\min\left\{\text{MDL}_{0s}+\text{MDL}_{1s},
l_s \right\} & \text{else}
\end{matrix}
\right..
\end{equation*}
In terms of the natural code, if $|s|=D$, then
$s$ is a leaf of the full depth-$D$ context tree,
and its natural code is empty; else $|s|<D$,
and the natural code requires $1$ bit to encode whether
$s\in\cl{S}$.
The symbols generated by $s$ are  encoded either
by retaining the children states (this requires
a coding length of $\text{MDL}_{0s}+\text{MDL}_{1s}$ bits),
or by pruning the children states and retaining
state $s$ with coding length $l_s$.
If $|s|=D$, then we do not process deeper contexts.
The CTP has $O(N/B)$ time complexity because the tree has $O(N/B)$ states.

\subsection{Phase~II} \label{subsec:basic:phaseII}

In Phase~II, PU~$b$ knows $\widehat{\cl{S}}$ and
$\{r_s\}_{s\in\widehat{\cl{S}}}$.
PU~$b$ encodes $x(b)$ sequentially;
for each symbol $x_i(b)$, it determines $G_i(b)$.
An $O(N/B)$ algorithm  for determining
$G_i(b)$ for all the symbols of $x(b)$ utilizing (\ref{eq:def:c_i(b)},\ref{eq:property:update_c_i}) is described by Baron~\cite{BaronThesis}.
After determining $G_i(b)$, the symbol $x_i(b)$
is encoded according to the probability assignment
$\widehat{p}(x_i(b))\triangleq\widehat{\theta}_{G_i(b)}^{x_i(b)}$
with an arithmetic encoder~\cite{Cover06}.
In order to have $O(N/B)$ time complexity and $O(1)$ expected coding redundancy per PU,
arithmetic coding is performed with $\log(N)$ bits of
precision~\cite{Cover06}, where we assume that the hardware architecture performs arithmetic with $\log(N)$ bits of precision in $O(1)$ time.

\subsection{Decoder} \label{subsec:basic:decoder}

The $B$ decoding blocks can be implemented on $B$ PUs.
Decoding block~$b$ decodes $x(b)$ sequentially;
for each symbol $x_i(b)$, it determines $G_i(b)$.
The same $O(N/B)$ algorithm used in Phase~II for
determining $G_i(b)$ for all the symbols of
$x(b)$ can be used in the $B$ decoding blocks.
After determining $G_i(b)$, the symbol $x_i(b)$
is decoded according to the probability assignment
$\widehat{p}(x_i(b))\triangleq\widehat{\theta}_{G_i(b)}^{x_i(b)}$
with an arithmetic decoder~\cite{Cover06} 
that has $O(N/B)$ time complexity.

\begin{theorem}~\cite{BaronThesis}
With computations performed with $\log(N)$ bits of precision defined as $O(1)$ time, the PTP-MDL encoder and decoder each require $O(N/B)$ time.
\end{theorem}

\section{Numerical Results} \label{sec:num_res}
\vspace*{-1mm}

This last section presents numerical results that compare the coding lengths of the parallel two-pass and naive parallel algorithms for different encoder settings and different numbers of parallel blocks. Two encoders are considered: MDL encoder (context tree pruning), and full depth Markov encoder (no pruning).

We test the average coding lengths over 2,000 repetitions
for signals of length $N = 10,000$ generated by the context tree source with $4$ states as described in Fig.~\ref{fig:tree}. The maximum context depth $D$ is set to be $5 \leq \log \left(N/B\right)$ for both MDL and Markov encoders. Hence the Markov encoder will run with $2^5 = 32$ states. For the MDL encoder, context tree pruning estimates the number of states $|\widehat{S}|$ to be around $4$, which is the number of states in the original source. Note that the coding length for a Bernoulli encoder, which uses $D = 0$,  is greater than the coding lengths of the other encoders, and hence is not included in our results.

\begin{figure}[t]
\vspace*{-5mm}
\hspace*{-5mm}
\includegraphics[angle=0,width=9cm]{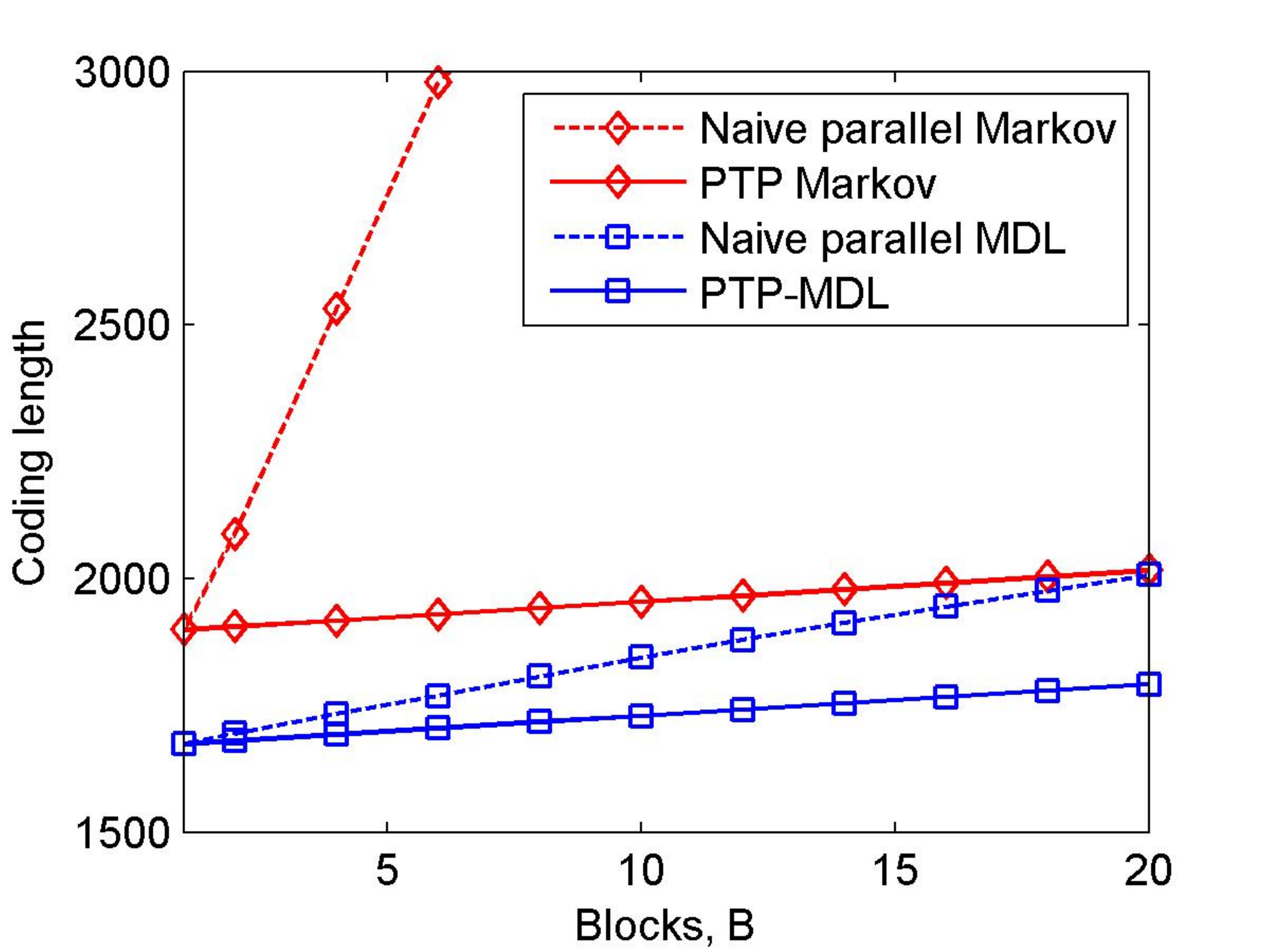}
\vspace*{-4mm}
\caption
{The performance of PTP and naive parallel compression algorithms using MDL and Markov encoders.}
\vspace*{-5mm}
\label{fig:plot}
\end{figure}

Fig.~\ref{fig:plot} shows our numerical results. It can be seen that PTP-MDL gives the best compression among the encoders, because the redundancy due to the source description $l^I_{\cl{S}}$ is higher for the Markov source than the MDL source due to the larger number of states.

Comparing the coding lengths for PTP-MDL and naive parallel compression, we can see that the rate of increase in coding length is higher for naive parallel compression. Both PTP-MDL and naive parallel compression suffer from the same coding redundancy, and redundancy due to unknown context for each block. However, the parameter redundancy due to source description is approximately $B$ times larger for naive parallel than for PTP-MDL. 

In summary, for context tree sources of depth $D \leq \log \left( N/B \right)$, PTP-MDL can compress data in $O(N/B)$ time while achieving a redundancy within $B\log(N/B)$  bits above Rissannen's lower bound on universal coding performance.

\section*{Acknowledgements}
\vspace*{-1mm}
This work was supported in part by the National Science Foundation under Grant CCF-1217749 and in part by the U.S. Army Research Office under Grant W911NF-04-D-0003. We thank Yoram Bresler for numerous discussions relating to this work;  Frans Willems for the arithmetic code implementation;
and Yanting Ma, Jin Tan, and Junan Zhu for their careful evaluation of the manuscript.

\bibliographystyle{IEEEbib}
\bibliography{cites}

\end{document}